\documentclass[twoside]{article}
\usepackage{amsmath}
\usepackage[psamsfonts]{amssymb}
\usepackage{cmmib57}
\newcommand{\bPf}{\par\vspace*{-4pt}\indent{\sc Proof.}\enskip}
\newcommand{\ePf}{\medskip}
\def\QED{\hskip0.1em\hfill\null\ \null\nobreak\hfill\kern3pt\vbox{\hrule\hbox
   {\vrule\kern1pt\vbox{\kern1.7pt\hbox{$\scriptscriptstyle{QED}$}
    \kern0.2pt}\kern1pt\vrule}\hrule}}

\def\END{\hskip0.1em\hfill\null\ \null\nobreak\hfill\kern3pt\vbox{\hrule\hbox
   {\vrule\kern1pt\vbox{\kern1.7pt\hbox{$\,\,\,\vspace{5pt}$}
    \kern0.2pt}\kern1pt\vrule}\hrule}}
\newtheorem{theorem}{Theorem}
\newtheorem{lemma}{Lemma}
\newtheorem{corollary}{Corollary}
\newtheorem{proposition}{Proposition}
\newtheorem{remark}{Remark}
\newtheorem{definition}{Definition}
\newtheorem{example}{Example}
\newcommand{\bCd}{\bEq\begin{CD}}
\newcommand{\eCd}{\end{CD}\eEq}
\newcommand{\bcd}{\beq\begin{CD}}
\newcommand{\ecd}{\end{CD}\eeq}
\newcommand{\ben}{\begin{enumerate}}
\newcommand{\een}{\end{enumerate}}
\newcommand{\bEq}{\begin{eqnarray}}
\newcommand{\eEq}{\end{eqnarray}}
\newcommand{\beq}{\begin{eqnarray*}}
\newcommand{\eeq}{\end{eqnarray*}}
\newcommand{\bDf}{\begin{definition}\em}
\newcommand{\eDf}{\end{definition}}
\newcommand{\bLm}{\begin{lemma}}
\newcommand{\eLm}{\end{lemma}}
\newcommand{\bPr}{\begin{proposition}}
\newcommand{\ePr}{\end{proposition}}
\newcommand{\bTh}{\begin{theorem}}
\newcommand{\eTh}{\end{theorem}}
\newcommand{\bCr}{\begin{corollary}}
\newcommand{\eCr}{\end{corollary}}
\newcommand{\bRm}{\begin{remark}\em}
\newcommand{\eRm}{\end{remark}}
\newcommand{\bEx}{\begin{example}\em}
\newcommand{\eEx}{\end{example}}
\newcommand{\C}{\mathbb{C}}

\newcommand{\ie}{{\em i.e$.$} }
\newcommand{\eg}{{\em e.g$.$} }
\newcommand{\A}{\forall}

\newcommand{\btht}{\boldsymbol{\tht}}

\newcommand{\cH}{\mathcal{H}}
\newcommand{\cI}{\mathcal{I}}

\newcommand{\bp}{\boldsymbol{p}}

\newcommand{\bz}{\boldsymbol{z}}

\newcommand{\bE}{\boldsymbol{E}}

\newcommand{\bG}{\boldsymbol{G}}

\newcommand{\bK}{\boldsymbol{K}}

\newcommand{\bP}{\boldsymbol{P}}

\newcommand{\bS}{\boldsymbol{S}}

\newcommand{\bV}{\boldsymbol{V}}

\newcommand{\bX}{\boldsymbol{X}}

\newcommand{\bZ}{\boldsymbol{Z}}

\newcommand{\sub}{\subset}

\newcommand{\wed}{\wedge}

\newcommand{\bet}{\beta}
\newcommand{\gam}{\gamma}
\newcommand{\del}{\delta}
\newcommand{\eps}{\epsilon}

\newcommand{\tht}{\theta}

\newcommand{\lam}{\lambda}

\newcommand{\ome}{\omega}
\newcommand{\Gam}{\Gamma}

\newcommand{\Ome}{\Omega}

\title{{\Large {\bf Towers with skeletons for the $(2+1)$-dimensional \\ 
continuous isotropic Heisenberg spin model}}}

\author{ Marcella Palese \\
{\footnotesize Department of Mathematics, University of Torino} \\ {\footnotesize via C. Alberto 10, I-10123 Torino, Italy } \\  {\footnotesize {\sc e-mail: marcella.palese@unito.it}}}

\date{}

\begin{document}

\maketitle

\begin{abstract}
We associate a tower with an infinitesimal algebraic skeleton to the $(2+1)$-dimensional (compact and noncompact) Heisenberg spin model. In particular, we construct the absolute parallelism defining the tower and the corresponding extension of the adjoint Lie algebra representation defining its skeleton.

\medskip


\end{abstract}

%---------------------------------------------------------------------------
\section{Introduction}
%--------------------------------------------------------------------------

The algebraic-geometric  approach to integrability of nonlinear systems is based on the {\em request of the 
existence of conservation laws} which leads to the existence of symmetries expressed in terms of {\em algebraic structures}.
The concept itself of integrability is a nontrivial matter and different definitions have been proposed in the literature on the subject, see \eg \cite{AbSe81}. We shall consider a concept of integrability 
as of having `enough' conservation laws to exaustively describe the dynamics.

It is a well known fact that the study of equations at least locally variational (\ie satisfying Helmholtz conditions) and thus arising as Euler--Lagrange equations of a (local) Lagrangian, enables one to precisely characterize conservation laws associated with symmetries of equations. 
From a physical point of view, in fact,  field equations appear to be a fundamental object, since they describe the changing of the field in base space; symmetries of equations are transformations of the space leaving invariant the description of such a change.
On the other hand the possibility of formulating a variational principle (\ie a principle of {\em stationary} action) - from which both changing of fields and associated conservation laws (\ie quantities not changing in the base space) could be obtained  - enables to keep account of both what ({\em and how}) changes and what ({\em and how}) is conserved. In the variational calculus perspective Euler-Lagrange field equations are `adjoint' to stationary principles up to conservation laws  \cite{FrPaWi12}.

The question is now what can we say when we do not know, in principle, if a variational formulation, even local, of a given nonlinear model, is possible.
It is clear that we would like to formulate a corresponding (maybe weaker) version of Helmholtz conditions of local variationality, being the Helmholtz morphism nothing but a quotient morphism of the exterior differential.
In previous papers, we proposed an algebraic-geometric formulation in terms of integrable towers with infinitesimal algebraic skeletons suitably associated with nonlinear models, see  \eg \cite{PaWi03,PaWi10,PaWi11}.
The tower is constructed in such  a way that both symmetries and conservation laws `along equations' can be deduced by an integrability condition.
In this paper we will consider algebraic structures also called `open' Lie algebra structures, in the sense that not all the commutators (\ie not all the Lie algebra structure constants) are determined,  found by Wahlquist and Estabrook 
for the study of integrability properties of nonlinear dispersive systems, and related with the existence of an infinite set of associated conservation laws generated by pseudopotentials \cite{WaEs75,EsWa76}. 
Such algebras are different from freely generated infinite-dimensional Lie algebras and the geometric interpretation of them was the object of various studies, se \eg 
\cite{PRS77,Es82,He76,Pal93,PaWi02,PaWi03,Pa05,PaWi10, PaWi11,PRS79} and references therein. 

In this note, we show that a tower with an infinitesimal algebraic skeleton can be associated with the $(2+1)$-dimensional continuous isotropic (compact and noncompact) Heisenberg spin model. In particular, we construct the absolute parallelism defining a tower, \ie a connection $1$-form  with values in an infinite dimensional algebra associated with the model, and the corresponding extension of the adjoint Lie algebra representation defining its skeleton.

%-----------------------------------------------------------------------------------
\section{The $(2+1)$-dimensional Heisenberg spin model}
%---------------------------------------------------------------------------------

In \cite {Pal93} algebraic properties of the $(1+1)$ dimensional (compact and noncompact) Heisenberg spin model, already well known to admit a Lax pair, were studied in both the direct and inverse prolongation structure procedure. In particular, the inverse prolongation (starting from the prolongation algebra) provided a whole family of spin models in $(1+1)$ dimension.

In this note, we shall investigate algebraic properties of a direct  extension in more than one spatial dimensions: the continuous isotropic (compact and noncompact) Heisenberg model in $(2+1)$ dimension given by
\bEq
(\Gam\bS)_t = \bS \times (\bS_{xx} +\bS_{yy})\,, \\ \nonumber
(\Gam\bS)\cdot \bS = \gam^2\,, \\
\Gam = \textstyle{diag} (1,1,\gam^2)\,, \qquad \gam^2= \pm1\,.
\eEq
Many variants of the model have been studied within different approaches;
however the integrability properties of the present model obtained as the direct extension in $(2+1)$ dimension of the $(1+1)$-dimensional Heisenberg spin field model are not yet completely understood \cite{Lak11}; in particular see also \cite{SLGR06,RGT00}. 

The construction of towers with skeletons associated with nonlinear field equations in $(2+1)$ dimension is highly non trivial \cite{Pa05,PaWi10,PaWi11}. We introduce some important concepts generalizing the concept of a homogeneous space and of a Cartan connection. To this aim let us represent this model by the following closed exterior differential system
\beq
&\btht_1 = d\bS \wed dy \wed dt - \bS_x dx\wed dy \wed dt\,, \\
&\btht_2 = d\bS \wed dx \wed dt + \bS_y dx\wed dy \wed dt\,, \\
&\btht_3 = d(\Gam\bS) \wed dy \wed dy + \bS \times (d\bS_x \wed dy \wed dt - d\bS_y dx \wed dt)\,, \\
&\bet_1 = d(\Gam\bS) \cdot \bS_x \wed dy \wed dt + (\Gam\bS) \cdot d\bS_x \wed dy \wed dt\,, \\
&\bet_2 = d(\Gam\bS) \cdot \bS_y \wed dx \wed dt + (\Gam\bS) \cdot d\bS_y \wed dx \wed dt \,.
\eeq

According to \cite{Mo93}, we  generalize the notion of homogeneous spaces by  defining an {\em algebraic skeleton} on a finite-dimensional vector space $\bV$ as a triple $(\bE,\bG,\rho)$, with 
$\bG$ a (possibly infinite-dimensional) Lie group, 
$\bE=\bV\oplus\mathfrak{g}$ is  a (possibly infinite-dimensional) vector space {\em not necessarily equipped with a Lie algebra structure}, 
$\mathfrak{g}$ is the Lie
algebra of $\bG$, and 
$\rho$ is a representation of $\bG$ on $\bE$ (infinitesimally of $\mathfrak{g}$ on $\bE$) such that it reduces to the adjoint representation of $\mathfrak{g}$ on itself.

Within our perspective, it is important to stress that the Lie algebra $\mathfrak{g}$ and the representation $\rho$ are the unknowns and that we want to formulate an approach which could enable us to determine them starting by the nonlinear model and an integrability condition for it. Once we know them, we can also construct $\bV$. 

Let us then introduce a manifold $\bP$ on which a Lie group $\bG$, with Lie algebra $\mathfrak{g}$, acts on the right; $\bP$ is a principal bundle $\bP\to\bZ\simeq\bP/\bG$. By construction, we have that $\bZ$ is a manifold of type
$\bV$, \ie $\A \bz\in \bZ$,  $T_{\bz} \bZ \simeq \bV$.
A tower $\bP(\bZ,\bG)$ on $\bZ$ with skeleton $(\bE, \bG,\rho)$ is an 
{\em  absolute parallelism} $\Ome$ on $\bP$ valued in $\bE$, invariant with respect to $\rho$ and reproducing elements of $\mathfrak{g}$ from the fundamental
vector fields induced on $\bP$. In general, the absolute parallelism  {\em does not}  define a Lie algebra homomorphism.

The representation $\rho$ defines a left action of $\bG$ on $\bE$ so that we can construct 
a bundle $\bP'= \bP\times_\rho \bE$, such that $V_{\bp'} \bP' \simeq\bE$.
This situation generalizes the standard construction of a principal bundle $\bP'= \bP\times_{\bG} \bK$, with $\bG$ a closed subgroup of $\bK$, or more precisely, if we  consider the action of $\bG$ on the Lie algebra $\mathfrak{k}$ of $\bK$,  $\bP'= \bP\times_{\bG} \mathfrak{k}$. The generalization here is that $\bE$ is a skeleton rather than a Lie algebra. 
We can think of a right action on $\bP'$ induced by $\rho$ in analogy with the homogeneous case; consequently, we consider a connection on $\bP'$ as a section of $J_1 \bP'\to\bP'$ invariant with respect to the induced right action. Accordingly,   $J_1 (\bP\times_\rho \bE)\to\bZ$ can be thought as the bundle of invariant connections with respect to $\rho$.

Let $\mathfrak{k}$ be a Lie algebra and $\mathfrak{g}$ a Lie subalgebra of
$\mathfrak{k}$. Let $\bG$ be a Lie group with Lie algebra $\mathfrak{g}$ and $\bP(\bZ, \bG)$ be a principal fiber bundle with structure group $\bG$ over a manifold $\bZ$ as above. 
A {\em Cartan connection} in $\bP$ of type
$(\mathfrak{k}, \bG)$ is 
a $1$--form $\Ome$ on $\bP$ with values in
$\mathfrak{k}$ such that
$\ome |_{T_{\bp} \bP}: T_{\bp} \bP\to \mathfrak{k}$ is an isomorphism $\forall \bp \in
\bP$, 
$R^{*}_{g}\ome=Ad(g)^{-1}\ome$ for $g\in \bG$ and reproducing elements of $\mathfrak{g}$ from the fundamental
vector fields induced on $\bP$.
Then it is clear that a Cartan connection
$(\bP, \bZ, \bG, \ome)$ of type $(\mathfrak{k}, \bG)$ is a special case of a tower on $\bZ$.

It is well known that in the homogeneous case, there is  a one-to-one correspondence between Cartan connections on $\bP$ and principal connections on $ \bP' $ with certain properties. 
A Cartan connection on $\bP$, that is a form $\ome$ on $\bP$ valued in $\mathfrak{g}$, 
can be spread as a %pseudotensorial
form %of the type $(Ad, \mathfrak{g}) $
%(KN, II.5) 
over the whole $\bP'$ ($\bP\sub\bP')$ as
$\ome_{(\bp',g)} = Ad(g^{-1} )\pi^{*}_{\bP}\ome + \pi^{*}_{\bG}\ome_{\bG}$
where $\pi_p : \bP \times \bG \to \bP$ and $\pi^{*}_{\bG}: \bP \times \bG \to \bG$ are canonical projections, and $\ome_{\bG}$ is the left Maurer-Cartan form on $\bG$;  $\ome$ can be extended to a form on $\bP \times \bG$ as a pull-up of a connection form of a principal connection on $ \bP' =  \bP \times_{\bK} \bG$.
Such a  construction gives rise to principal connections on $ \bP' $, the
horizontal bundle of which does not intersect the tangent bundle of $\bP$ viewed as the subbundle 
of $ \bP' $; conversely the pull-back under the canonical inclusion $i : \bP \to \bP \times_{\bK} \bG$ of the connection form of a principal connection, which satisfies such a condition, is a Cartan connection on $\bP$ with values in $\mathfrak{g}$. 

Let us come back into the situation of a skeleton instead of a homogeneous space:  we generalize this construction 
by requiring  that $\bP$ is a manifold of type $\bE$, \ie $T_{\bp}\bP\simeq\bE$, and $V_{\bp'} \bP' \simeq T_{\bp}\bP$.
A tower can be seen then as an invariant connection with respect to $\rho$ on $\bP'$ satisfying the latter condition.
Under this perspective, $\bP'= \bP\times_\rho \bE$ can be considered a sort of a gauge-natural bundle.
and a tower $\bP$ with skeleton  $\bE$ could be considered as a section of a subbundle of $\bP'$.
Let us now express locally the forms of the  induced absolute parallelism
\beq
& \Ome^k = H^k (\bS, \bS_x, \bS_y; \xi)dx \wed dy + F^k (\bS, \bS_x, \bS_y; \xi) dy \wed dt +
\\
&+ \, G^k (\bS, \bS_x, \bS_y; \xi)dx \wed dt + (A^k_m dx+B^k_m dy+C^k_m dt)\wed  d\xi^m \nonumber \,,
\eeq
where $F^k, G^k, H^k$ are functions to be determined modulo $\btht_1\,, \btht_2\,, \btht_3 \,,\bet_1\,,\bet_2$ and $A, B, C$ are invertible matrices
\footnote{Notice that the absolute parallelism is actually given by
$
\ome^{m}=\Gam^{m}_{1}dx+\Gam^{m}_{2}dy+\Gam^{m}_{3}dt+d\xi^{m}
$,
so that $H^{k}= \Gam^{m}_{1}B^{k}_{m}-\Gam^{m}_{2}A^{k}_{m}$, $F^{k}= \Gam^{m}_{2}\del^{k}_{m}-\Gam^{m}_{3}B^{k}_{m}$,
$G^{k}=\Gam^{m}_{1}\del^{k}_{m}-\Gam^{m}_{3}A^{k}_{m}$. }.

Let $\cI$ be the ideal generated by the forms 
${\btht}_{i},\bet_{l}$, $i=1,2,3$, $l=1,2$ and
$\Omega^{k}$, $k=1,\ldots,N$.  We say that $\cI$ is closed if 
$d\Omega^{k}\in \cI$
$({\btht}_{i},\bet_{l},\Omega^{k})$.

The integrability condition $d\Ome^k =0$ $(\textstyle{mod} \, \cI)$  yields
\beq
H^k_{\bS_x}= 0\,, \quad H^k_{\bS_y}=0\,, \quad \Rightarrow H^k= H^k(\bS,\xi) \,,\\
F^k_{\bS_x}= - (\Gam H^k_{\bS})\times \bS \,, \quad  F^k_{\bS_y}=0 \Rightarrow F^k= F^k(\bS, \bS_x,\xi)\,, \\
G^k_{\bS_y}= (\Gam H^k_{\bS})\times \bS \,,  \quad  G^k_{\bS_x}=0 \Rightarrow G^k= G^k(\bS, \bS_y,\xi)\,,
\eeq
and a further important constraint:
\beq
& (F^k_{\bS}\cdot\bS_x - G^k_{\bS}\cdot\bS_y)dx \wed dy \wed dt +
H^k_{\xi^m} d\xi^m \wed dx \wed dy + \\
& +  F^k_{\xi^m} d\xi^m \wed dy \wed dt + G^k_{\xi^m} d\xi^m \wed dx \wed dt =0 \,.
\eeq
Without loosing of generality, we can assume $C^m_l =\del^m_l $ so that, substituting 
\beq
&dt\wed d\xi^m= \Ome^m -   H^m dx \wed dy - F^m dy \wed dt +\\ 
& - \,G^m dx \wed dt - A^m_l dx \wed d\xi^l -B^m_l dy\wed d\xi^l\,
\eeq
gives us the fundamental constraints
\bEq
F^k_{\bS}\cdot\bS_x - G^k_{\bS}\cdot\bS_y + [G,F]^k=0\,,\label{constraint} 
\\
H^k_{\xi^l} +F^k_{\xi^m}A^m_l - G^k_{\xi^m}B^m_l =  0\,.
\eEq
From the second equation we can easily infer that $[A,B]=0$; furthermore, the  functions $H$, $G$ and $F$ are related as follows (to simplify the notation, we omit the superscripts)
\bEq\label{constr}
[G, F] = [\bar{B} H, \bar{B} F]\,.
\eEq
%or alternatively $\bar{B}[H, F]= [\bar{B} H, \bar{B} F]$, which implies $\bar{B}[H, F]=[G, F] $.
Constraints above can be further recasted as follows
\beq
& F = - (\Gam H_{\bS})\times \bS\cdot \bS_x + K(\bS;\xi)\,, \\
& G =  (\Gam H_{\bS})\times \bS\cdot \bS_y + \bar{K}(\bS;\xi)\,,
\eeq
thus, in particular, $F_{\bS}\cdot \bS_x = K_{\bS}\cdot \bS_x $ and $G_{\bS} \cdot \bS_y = \bar{K}_{\bS}\cdot \bS_y$.
By substitution in \eqref{constraint} we thus obtain $
H=\bX(\xi)\cdot\bS+Y(\xi)$, where $\bX =(X_1, X_2, X_3)$ (therefore $H_{\bS}= \bX$), so that 
$F$ $=$ $ -(\Gam\bX)\times \bS \cdot\bS_x$ $+$ $ K(\bS;\xi)$, $G$ $=$ $(\Gam\bX)\times \bS \cdot\bS_y $ $+$ $ \bar{K}(\bS;\xi)$.
We can also verify that the main constraint also implies $K_{\bS}=0$ thus $K=K(\xi)$. On the other hand it also implies $\bar{K} = - S_1[X_2, X_3] + S_2[X_1, X_3] - \gam^2 S_3[X_1, X_2] + Z(\xi)$.
We can summarize that by
\beq
& H= \bX\cdot\bS +Y\,, \\
& G = (\Gam \bX\times\bS)\cdot \bS_y -  S_1[X_2, X_3] + S_2[X_1, X_3] - \gam^2 S_3[X_1, X_2] +Z\,, \\
& F = - (\Gam \bX\times\bS)\cdot \bS_x + K\,.
\eeq
To uniform the notation, let us put $Y=X_4$ and $Z=X_5$ (notice that $K$ is is related with them according with the relationship \eqref{constr}); by substitution and comparing terms of the same monomials, after long but simple algebraic manipulations, we obtain an  algebra structure $\bE$ generated by $X_1$, $X_2$, $X_3$, $X_4$, $X_5$:
\bEq
[ X_1 , X_2 ] = 0 \,, [ X_1 , X_3] = 0 \,, [ X_1, X_4 ]  = X_6 \,,  [ X_1, X_5 ] = X_7 \,, \label{open}
\eEq
\beq
[ X_2 , X_3 ] = 0 \,, [ X_2 , X_4 ] = X_8 \,, [ X_2, X_5 ] = X_9 \,,
\eeq 
\beq
[X_3 , X_4] =X_{10}\,, [ X_3, X_5] = X_{11} \,, [ X_4, X_5 ] = X_{12}\,,
\eeq
\beq
\ldots \qquad  \ldots \qquad  \ldots \qquad  \ldots \qquad 
\eeq
By requiring the Jacobi identity to hold true, it is easy to verify that the structure does not close as a Lie algebra, since always new generators have to be introduced to name the commutators which are unknown.
This open structure is infinite-dimensional and identifies an infinite dimensional vector space $\bE$. It is different from a freely generated Lie algebra insomuch as there are some relations among some of the commutators.
It  can be provided of the 
structure of an infinitesimal algebraic skeleton on a finite dimensional space $\bV$. We define a Lie algebra $\mathfrak{g}$ acting on $\bE$ by the representation $\rho$ obtained by means of the request of integrability for the {\em absolute parallelism} of a tower on $\bZ$, with skeleton  $(\bE, \bV, \mathfrak{g})$. 

%-----------------------------------------------------------------------------------------------------------------
\subsection{Skeletons homomorphic with a finite dimensional quotient Lie algebra}
%-----------------------------------------------------------------------------------------------------------------

We can identify some special subalgebras and the corresponding extension of the adjoint Lie algebra representation defining skeletons as follows.

\begin{enumerate}
\item Put $X_1= 0 \,, X_2=0$, \ie $\bX=(0, 0, X_3)$. We then also have $X_6=  X_7= X_8= X_9=0$.

The resulting structure 
\beq
& [X_3 , X_4] =X_{10}\,, [ X_3, X_5] = X_{11} \,, [ X_4, X_5 ] = X_{12}\,, 
\\
& [X_3 , X_{12}] = [X_4 ,X_{11}] - [X_5 ,X_{10}] \,, 
\\
& \ldots \qquad  \ldots \qquad  \ldots \qquad  \ldots \qquad 
\eeq
identifies a tower with skeleton defined by
\beq
& H = S_3 X_3  + X_4 \,,
\\
& G = \gam^2(S_1  S_{2y} - S_2  S_{1y}  )X_3 + X_5\,,
\\
& F = \gam^2(S_2  S_{1x} - S_1 S_{2x}  )X_3 + K\,; 
\eeq
notice that $[X_5, K] =[\bar{B}X_4,\bar{B}K]$, so that, without loosing generality, we can take $K=X_{12}$.

\item Put $X_1= 0 \,, X_3=0$, \ie $\bX=(0,  X_2, 0)$. We then also have $X_6= X_7= X_{10}= X_{11}=0$.

The resulting structure 
\beq
& [X_2 , X_4] =X_{8}\,, [ X_2, X_5] = X_{9} \,, [ X_4, X_5 ] = X_{12}\,, 
\\
& [X_2 , X_{12}] = [X_4 ,X_{9}] - [X_5 ,X_{8}] \,, 
\\
& \ldots \qquad  \ldots \qquad  \ldots \qquad  \ldots \qquad 
\eeq
identifies a tower with skeleton defined by
\beq
& H = S_2 X_2  + X_4\,,
\\
& G = \gam^2(  S_1  S_{3y} -S_3  S_{1y}  )X_2 + X_5\,,
\\
& F = \gam^2(S_3 S_{1x}  - S_1  S_{3x} )X_2 + X_{12}\,.
\eeq

\item Put $X_2= 0 \,, X_3=0$, \ie $\bX=(X_1, 0,  0)$. We then also have $X_8=  X_9=  X_{10}=  X_{11}=0$.

The resulting structure 
\beq
& [X_1 , X_4] =X_{6}\,, [ X_1, X_5] = X_{7} \,, [ X_4, X_5 ] = X_{12}\,, 
\\
& [X_1 , X_{12}] = [X_4 ,X_{7}] - [X_5 ,X_{6}] \,, 
\\
& \ldots \qquad  \ldots \qquad  \ldots \qquad  \ldots \qquad 
\eeq
identifies a tower with skeleton defined by
\beq
& H = S_1 X_1  + X_4\,,
\\
& G = \gam^2(S_2  S_{3y} - S_3  S_{2y}  )X_1 + X_5\,,
\\
& F= \gam^2(S_3  S_{2x} - S_2 S_{3x}  )X_1 + X_{12}\,.
\eeq
\end{enumerate}

Such skeletons just differ for a renaming of some of the elements of the algebra \eqref{open} and then identify the same algebraic structure.
The vector space $\bV$ can be defined as the kernel of an homomorphism between the infinite dimensional algebra \eqref{open} and a finite-dimensional Lie algebra $\mathfrak{g} = \mathfrak{s}\mathfrak{l}(2, \C)$. 

\bPr
There exists an homomorphism $\cH$ between the algebraic structure defined above (and thus also between $\bE$) and the $\mathfrak{s}\mathfrak{l}(2, \C)$ Lie algebra.
\ePr
\bPf
The homomorphism $\cH$ is defined by the closing conditions
\beq
X_{10} = 2i\lam X_5 \,,  \qquad X_{11} = - 2i\lam X_4 \,, \qquad  X_{12} = 2i\lam X_3 \,,
\eeq
(analogously $X_{8} = 2i\lam X_5 \,, X_{9} = -2i \lam X_4 \,, X_{12} = 2i\lam X_2$, or
$X_{6} = 2i\lam X_5 \,, X_{7} = - 2i\lam X_4 \,, X_{12} = 2i\lam X_1$, for the other cases); here $\lam$ is a parameter.
The $\mathfrak{s}\mathfrak{l}(2, \C)$ Lie algebra is then given by 
$[X_i, X_j ] = 2i\lam \eps^{ijk} X_k$, with $i,j, k= 3,4,5$ (resp. $i,j, k= 2,4,5$ and $i,j, k= 1,4,5$).
\ePf

Representations of such a quotient Lie algebra $\mathfrak{s}\mathfrak{l}(2, \C)$ provide conservation laws in the form of linear spectral problems associated with the continuous isotropic (compact and noncompact) Heisenberg model in $(2+1)$ dimensions starting from the towers constructed in $(i)-(iii)$; in fact, each tower provides one of the components in which the model can be decomposed with respect to a base. Notice also that the same algebra can be obtained directly from $\bE$ by setting $ X_1 =X_2 =X_3 = - \frac{i}{2\lam}X_{12}$.

%---------------------------------------
\section*{Acknowledgements}
%---------------------------------------
Research supported by Department of Mathematics, University of Torino, project  {\em Metodi Geometrici in Fisica Matematica e Applicazioni $(2011)$}. The author thanks also the partial support of the Czech grant GA$201/09/0981$ {\em Globalni analyza a geometrie fibrovanych prostoru}, as well as  the University of Brno and the University of Ostrava, where part of this work has been written. 
Thanks are due to E. Winterroth for useful remarks.

Dedicated to Hartwig.

%---------------------------------------
\section*{References}
%---------------------------------------

\end{document}